\documentclass[journal,twocolumn]{IEEEtran}
\ifCLASSINFOpdf
  \usepackage[pdftex]{graphicx}
  \graphicspath{{../pdf/}{../jpeg/}}
  \DeclareGraphicsExtensions{.pdf,.jpeg,.png}
\else
  \usepackage[dvips]{graphicx}
  \graphicspath{{../eps/}}
  \DeclareGraphicsExtensions{.eps}
\fi
\usepackage{color}
\usepackage{epsfig}
\usepackage{booktabs}
\usepackage{epstopdf}
\usepackage{amssymb}
\usepackage{mathtools}
\usepackage{algorithm}
\usepackage{algpseudocode}
\usepackage{array}
\usepackage{mdwmath}
\usepackage{mdwtab}
\usepackage{bm}
\usepackage{mathrsfs}
\usepackage{epsfig}
\usepackage{subfigure}
\usepackage{amsmath}
\usepackage{mathrsfs}
\usepackage{cite}
\usepackage{url}
\usepackage{tikz}
\usetikzlibrary{quantikz}

\DeclareGraphicsExtensions{pst,eps}
\graphicspath{{./fig/}}

\newtheorem{lemma}{Lemma}

\newcommand{\qQ}{{\bf Q}}
\newcommand{\qh}{{\bf h}}
\newcommand{\qJ}{{\bf J}}
\newcommand{\qa}{{\bf a}}
\newcommand{\qb}{{\bf b}}
\newcommand{\qH}{{\bf H}}
\newcommand{\qx}{{\bf x}}
\newcommand{\qs}{{\bf s}}
\newcommand{\qR}{{\bf R}}

\newcommand{\qg}{\bm{g}}
\newcommand{\qf}{\bm{f}}
\newcommand{\qone}{\bm{1}}
\graphicspath{{fig/}}
\newcommand{\be}{\begin{equation}} \newcommand{\ee}{\end{equation}}
\newcommand{\bea}{\begin{eqnarray}} \newcommand{\eea}{\end{eqnarray}}

\begin{document}

\title{Constrained Higher-Order Binary Optimization for  Wireless Communications Systems Using Ising Machines}
\author{Gan Zheng,~\IEEEmembership{Fellow,~IEEE,} and Ioannis Krikidis,~\IEEEmembership{Fellow,~IEEE}
}

\maketitle
\vspace{-2cm}
\begin{abstract}
 This paper develops an algorithmic  solution  using Ising machines  to solve large-scale higher-order binary optimization (HOBO) problems with inequality constraints for resource optimization in wireless communications systems. Quadratic  unconstrained binary optimization (QUBO) aims to solve a special category of these problems  widely encountered in engineering and science. To solve QUBO instances, specialized Ising machines have been designed, while sophisticated quantum annealing algorithm and quantum-inspired classical heuristics have been developed. However, the application of QUBO in wireless communications has limited practical interest mainly due to the complexity of resource optimization problems which are often characterized by high-order polynomial terms and strict inequality constraints. To overcome these bottlenecks and take advantage of recent advancements in Ising machines, in this paper, we propose an iterative algorithmic  solution   to solve HOBO problems, which is based on the augmented Lagrangian method to handle constraints.  Specifically, Taylor expansion is employed to approximate higher-order polynomials to quadratic  ones in the augmented Lagrangian function, which enables the solution of a single QUBO problem at each iteration without auxiliary variables. As an illustrative case study, we consider the problem of phase optimization in a  simultaneous wireless information and power transfer system, where a reconfigurable intelligent surface with $1$-bit phase resolution is used to facilitate information/energy transfer.  Simulation results verify that the proposed algorithm  achieves satisfactory performance and outperforms heuristic benchmark schemes.
\end{abstract}


\section{Introduction}
Binary optimization has broad applications in signal processing, machine learning, computer vision, and telecommunication engineering \cite{qubo}. In wireless communications systems, problems of binary optimization range from signal detection, channel decoding,  multiuser detection, power control to frequency assignment, subcarrier allocation, beam assignment in massive multiple-input multiple-output (MIMO), task offloading in mobile-edge computing, etc. However, binary optimization belongs to the NP-hard combinatorial problems and is difficult to solve especially when the problem's scale is large. Traditionally, binary optimization problems are solved by using heuristic  algorithms such as genetic algorithms \cite{GA}, simulated annealing \cite{SA0}, and branch and bound techniques \cite{bb}.
 In the sixth-generation (6G) wireless communication system, the rapid proliferation of communication devices—including antennas, users, and reconfigurable intelligent surfaces (RISs)—along with new functionalities such as computing and sensing, gives rise to large-scale optimization problems \cite{largescaleopt}\cite{beamselection}. These challenges are increasingly difficult for heuristic algorithms to handle effectively.

 Recently,  a new promising alternative approach  to address complex NP-hard combinatorial resource optimization problems in wireless communications has emerged.
This approach formulates a combinatorial optimization problem as a quadratic unconstrained binary optimization (QUBO) problem or Ising model and then solves the QUBO problem using Ising machines. Ising machines are purposely built physical machines that aim to find the ground states of an Ising model and therefore solve the equivalent
QUBO problems. Various Ising machines or QUBO solvers have been developed in recent years which can be categorized as quantum machines and classical machines.

 A typical quantum Ising machine is quantum annealer (QA). QA  uses the principles of the Adiabatic evolution \cite{MCG} as long as the problem  can be formulated in the format of an Ising model. D-WAVE is a commercial analogue quantum device that implements QA by using complex superconducting integrated circuits \cite{WAVE} and has been widely used in wireless communications research. For instance, QA  has been used to optimize nonlinear multiuser MIMO vector precoding in \cite{Annealing-MIMO} and this work has been further improved in  \cite{QA-WCL} by introducing appropriate preprocessing  called lattice reduction. Recent studies  have extended the use of  QA to optimize the analogue pre/post-coding vector design for a MIMO system with $1$-bit phase shifters \cite{KRI3} and  $1$-bit digital  complex-valued  pre/post coding vectors in \cite{KRI4}.

 Classical Ising machines are specially built hardware devices to solve QUBO problems inspired by quantum mechanics. They include Toshiba's Simulated
Bifurcation Machine \cite{SB}, Fujitsu’s Digital Annealer (DA) \cite{digital-annealer} and   Amplify Annealing Engine (AE) \cite{amplify}. These  algorithms have demonstrated high efficiency and accuracy. For instance, the work in \cite{inspired-compare} studies Max-Cut problems up to $20,000$ nodes, and shows that several classical Ising machines  outperform classical heuristics and QA in terms of time-to-solution performance. In another example, it was shown that  simulated bifurcation computes the optimal time and frequency allocation of $20$ terminals in less than $0.5$ ms \cite{SB-URLLC}. Motivated by the above results, in this paper, we focus on classical Ising machines  and especially Toshiba's simulated bifurcation   to demonstrate the efficiency of the proposed algorithmic  solution.

However, there are two main limitations of existing Ising machines: they cannot handle general constraints or polynomials of higher order than quadratic ones.  Many optimization problems in wireless communications have strict equality and  inequality resource constraints  and either the objective function or the constraints may involve high-order polynomials. Due to this observation, the practical application of the powerful Ising machines in the design of wireless communications systems is still limited.  On the one hand, to deal with inequality constraints (more challenging than equality ones) while still making use of Ising machines, an algorithm based on the alternating direction method of multipliers was designed and tested for quadratic knapsack problems in \cite{constraint-ADMM}. An efficient augmented Lagrangian approach to deal with multiple inequality constraints for solving the set cover problem was designed in \cite{Djidjev}. In a more recent work \cite{unbalanced}, an unbalanced penalization function was proposed to convert inequality constraints into the QUBO form, and tested for practical combinatorial problems such as the traveling salesman problem, the bin packing problem, and the knapsack problem.
On the other hand, the mainstream approach for solving high-order optimization problems using Ising machines is quadratization \cite{quadratization}, which involves introducing auxiliary variables to convert higher-order polynomial terms into quadratic or linear ones. However, this method significantly increases the number of variables, thereby introducing additional challenges for Ising solvers.

 To summarize, existing  traditional numerical algorithms cannot deal with the large scale of binary optimization while Ising machines  cannot handle higher-order polynomials or inequality constraints.  Distinct from existing works, the novelty of this paper lies in addressing both inequality constraints and higher-order polynomial objectives by designing an  algorithm capable of solving large-scale, constrained, high-order binary optimization problems. This advancement enables practical applications in wireless communications, which were previously not tractable with existing methods.
Our main contributions are summarized as follows:
\begin{itemize}
  \item  We propose an  algorithm for solving large-scale, higher-order constrained binary optimization problems by leveraging the computational power of Ising machines for QUBO. Our approach introduces a modified augmented Lagrangian framework capable of effectively handling inequality constraints. Additionally, we develop a solution strategy based on a second-order Taylor approximation to address the higher-order terms that arise within the augmented Lagrangian formulation. This dual strategy enables our method to overcome two major limitations of existing approaches: the inability to manage inequality constraints and the inefficiency in dealing with higher-order polynomial objectives.

       The proposed method has significant interest as it provides a   hybrid   solution  that leverages the power of both classical and quantum Ising machines to address large-scale constrained and higher-order binary optimization problems.

  \item As an indicative case study, we apply the proposed algorithm to optimize the phase shifts of a   RIS-assisted simultaneous wireless information and power transfer (SWIPT) system. Specifically, we study the case where the quality of the received signal at the information receiver is maximized, while satisfying the minimum energy requirement following a practical nonlinear energy-harvesting model.  This is a difficult nonlinear constrained binary optimization problem for which no efficient solution is known.  Numerical results show that the proposed algorithm   achieves satisfactory performance for such a RIS-assisted system.
\end{itemize}

The remainder of this paper is organized as follows. Section II introduces the QUBO problem formulation, various classical Ising machines available, as well as the general augmented Lagrangian algorithm for solving constrained optimization.  Section III presents the proposed modified  augmented Lagrangian algorithm for solving inequality constrained HOBO. Section IV gives an application example in which the binary phase shifts  of a RIS are optimized to balance the performance of information and energy transfer in a basic SWIPT setup. Simulation results and conclusions are presented in Section V and Section VI, respectively.

{\em Notions:} All boldface letters indicate vectors (lower case) or matrices (upper case).  The superscripts   $(\cdot)^T$ and  $(\cdot)^{\dag}$ denote the transpose and conjugate transpose of a matrix, respectively. In addition, $\qx\in \mathcal{C}^{N\times 1}$ or $\qx\in \mathcal{R}^{N\times 1}$ denotes that $\qx$ is an $N\times 1$ complex or real vector; $\mbox{sign}(\cdot)$ return the sign of the input variable; $\mbox{Re}(\cdot)$ returns the real part of the input variable.  $\mbox{Diag}(\qx)$ returns a diagonal matrix with  the elements of the vector $\qx$ on the main diagonal.

\section{QUBO and Constrained Optimization}
In this section, we first introduce the QUBO problem and present the classical Ising machines as QUBO solvers. In addition, we provide the mathematical formulation of the constrained binary optimization problem and highlight the main classical numerical algorithms to solve it.

\subsection{QUBO problem formulation}
Consider an $N$-dimensional binary variable vector $\qx$ where $x_i\in \{0,1\}, \forall i=1, \cdots, N$. The objective function of a standard QUBO problem is defined as
\be\label{eqn:QUBO_Boolean}
    f(\qx) \triangleq \qx^T \qQ \qx = \sum_{i>j} Q_{i,j} x_i x_j + \sum_{i=1}^N Q_{i,i} x_i, \;\;\;x_i \in \{0,1\}, \forall i,
\ee
where $\qQ$ is a real lower triangular matrix. Note that $x^2_i=x_i$, so the second term also represents a quadratic term.

We call the above formulation as a Boolean model. If the variables belong to the spin set $\{-1,1\}$, the problem is called an Ising model and is formulated as
\be\label{eqn:QUBO_Ising}
    f(\qs) \triangleq \qs^T \qJ \qs+ \qa^T \qs = \sum_{i>j} J_{i,j} s_i s_j + \sum_{i=1}^N a_{i} s_i, \;\;\;s_i \in \{-1,1\}, \forall i,
\ee
where $\qJ$ is a real lower triangular matrix and $\qa$ is a real vector. For clarify, we have used the spin variable vector $\qs$ to differentiate  it from the Boolean model. Note that since $s_i^2=1$, there is no quadratic term corresponding to a single variable, {\it i.e.,} the diagonal elements of the matrix $\qJ$ are all zero.

Clearly the Boolean and Ising variables are equivalent by a simple linear transformation, and therefore we can convert the Ising model to a Boolean model equivalently and vice versa. It is worth noting that in several cases, this transformation is necessary since some QUBO solvers can only handle one of the two types of binary problems (usually Boolean).  For the sake of presentation, we provide  an  example related to the conversion of the Ising model to the Boolean model.

Suppose $\qs = 2\qx-1$, then the objective function in \eqref{eqn:QUBO_Ising} can be rewritten as
\bea
    f(\qs) &= &f(2\qx-1)\notag\\
     & =&  \sum_{i>j} J_{i,j} (2x_i-1) (2x_j-1) + \sum_{i=1}^N  a_{i} (2x_i-1)\notag\\
    &= & \qx^T \bar\qQ \qx + \qone^T \qJ \qone - \qa^T \qone,
\eea
where  $\qone$ is an all-ones vector of appropriate dimension.  The new matrix $\bar \qQ$ is defined as
\be
 \bar Q_{i,j}=
\begin{cases}
    4J_{i,j}\,& \text{if } i>j\\
    b_i,              & \text{if } i=j\\
    0, & \text{otherwise,}
\end{cases}
\ee
where $\qb$ is defined as
\be
    \qb = 2\qa - 2(\qJ + \qJ^T)\qone.
\ee

\subsection{Classical Ising Machines}
 In recent years, there has been rapid development in classical Ising machines that aim to find the absolute or approximate ground state of the Ising model inspired by quantum annealing.  In this subsection, we briefly introduce some representative examples.

Quantum bifurcation machine (QbM) solves quantum adiabatic optimization based on quantum bifurcations of Kerr-nonlinear parametric oscillators.  A classical machine which is inspired by QbM, called `Simulated Bifurcation Machine (\textit{SBM})',     has been   shown to exhibit high performance for large-scale combinatorial optimization problems  \cite{SB}. SBM numerically simulates the adiabatic evolution of classical nonlinear Hamiltonian systems exhibiting bifurcation phenomena, where the two branches of the bifurcation in each nonlinear oscillator correspond to the two states of each Ising spin. In this subsection, we only introduce the discrete SBM (dSB), an efficient version of SBM \cite{SB3}.

To start with, suppose the Ising problem to find the spin configuration that minimizes the Ising energy  defined as
\be\label{eqn:E}
    E \triangleq \sum_{i=1}^N \sum_{j=1}^N J_{i,j} x_i x_j.
\ee
Note that we have ignored the linear terms because by introducing
ancillary spin variables, the linear terms can be incorporated in \eqref{eqn:E}. Here we slightly abuse of notation and still use $\{x_i\}$ to denote the spin variables.

To solve \eqref{eqn:E}, a new quantum adiabatic optimization that uses a network of Kerr-nonlinear parametric oscillators (KPOs) has   been proposed in \cite{oscillator}.  It was found later that the classical system corresponding to QbM can also work as
an Ising machine to  find good approximate solutions with high probability. This observation is significant in the sense that
we do not need to build a real quantum machine, but we can efficiently emulate such a system  by using classical computers.

By simplifying and modifying the Hamiltonian equations of QbM, we can obtain the following Hamiltonian equations
for dSB:
\bea
\dot{x_i}&=&\frac{\partial H_{dSB}}{\partial y_i}=b_0 y_i,\label{eqn:x}\\
\dot{y_i}&=&-\frac{\partial H_{dSB}}{\partial x_i}=-\left[b_0-b(t)\right]x_i \notag \\ && - \xi_0\sum_{j=1}^N J_{i,j} \mbox{sign}(x_j),\\ \label{eqn:y}
  H_{dSB}&=& \frac{b_0}{2}  \sum_{i=1}^{N} y_i^2 + V_{dSB},\\
  V_{dSB}&=& \begin{cases}
    \frac{b_0-b(t)}{2}\sum_{i=1}^{N} x_i^2   -\xi_0 \\  \sum_{i=1}^N\sum_{j=1}^N J_{i,j} x_i\mbox{sign}(x_j), \\\text{if $|x_i|<1, \forall x_i$}, \\
     \infty, & \text{otherwise,}
\end{cases}
\eea
where  $x_i$ and $y_i$ are, respectively, the position and momentum (an auxiliary variable) of a
particle corresponding to the $i-$th spin,  $b(t)$ is a control parameter increased from zero, $b_0$ and $\xi_0$ are positive
constants, $V_{dSB}$ is the potential energy in dSB, and  $\mbox{sign}(x_i)$ corresponds to the solution of the spin $x_i$.

By using the above new formulation of the Hamiltonian and the Euler method, we have the following update rule for $x_i$ and $y_i$, {\it i.e.,}
\bea
    x_i(t_{k+1})& =& x_i(t_{k}) + b_0 \delta_t y_i(t_{k+1}), \label{eq:euler:x}\\
    y_i(t_{k+1})& = &  y_i(t_{k}) + \delta_t (-\left[b_0-b(t)\right]x_i(t_{k}) \notag \\
    &- & \xi_0\sum_{j=1}^N J_{i,j} \mbox{sign}(x_j(t_{k}))),\label{eq:euler:y}
\eea
where $\delta_t$ is the time step and $t_k$ is the discretized time satisfying $t_0 = 0$
and $t_{k + 1} = t_k + \delta_t$. After updating $x_i$, if the condition $|x_i| > 1$ holds, $x_i$ is replaced  with $\mbox{sign}(x_i)$ and  $y_i$ is set to $0$.

There are several advantages with the above simplified classical Hamiltonian system. First, it allows  treatment of
large-scale problems with dense connectivity, which are challenging for QA; second, the simplicity of the explicit  Euler updates \eqref{eq:euler:x}  and \eqref{eq:euler:y}  are crucial for enabling hardwiring of the dSB algorithm with efficient custom circuits. We also note that the expressions in \eqref{eq:euler:x}  and \eqref{eq:euler:y}  allow updating variables simultaneously at each time step, unlike simulated annealing that requires one-by-one updating to guarantee convergence. By exploiting this parallel computation structure, the dSB algorithm can be implemented by using various technologies, {\it e.g.,} FPGA, GPU and ASIC, to significantly accelerate its computations.

It is also worth mentioning that there is high-performance classical QUBO solver called  Fixstars Amplify Annealing Engine (Amplify AE) \cite{amplify}. Amplify AE is GPU-based Ising machine that can process large-scale problems consisting of more than $100,000$ bits at a high speed. It is based on  parallelized simulated annealing and uses multi-level parallel processing on multiple GPUs to find optimal solutions; for some problems, it is shown to outperform QA   \cite{amplify-benchmark}.

\subsection{Binary Optimization with equality and inequality constraints}
 A standard constrained binary optimization problem is formulated as
 \bea\label{prob0}
    \min_{\qx\in\{0, 1\} \mbox{~or~} \{-1,1\}} && f(\qx) \notag \\
    \mbox{s.t.} && h(\qx)=0,\notag\\
                && g(\qx)\le 0,
 \eea
where $f(\qx)$ is the objective function, $h(\qx)$ is the equality constraint function and $g(\qx)$ is the inequality constraint function. We do not impose any requirement on the convexity but assume all functions are polynomials that can have an order higher than two. The optimization variable  $\qx$ can be either a Boolean variable $\{0,1\}$ or an Ising spin variable $\{-1,1\}$.

 Problem  \eqref{prob0} is clearly NP-hard and cannot be directly solved using existing QUBO solvers, due to the presence of constraints and potentially higher-order polynomial terms. The treatment of higher-order optimization is deferred to Section III. In the remainder of this section, we review two existing methods for handling constraints, which are described below.
\begin{itemize}
    \item Penalty method is a straightforward way to convert the equality constraints to additional terms into the objective function. Usually the equality constraint is easier to solve, and its penalty term is represented as $\mu h^2(\qx)$ where $\mu$ is a large penalty factor.

        For the inequality constraint, it is more complicated. One method is to use a modified penalty term like $\mu(\max(0,g(\qx))^2$.
         Then the original constrained problem can be cast into the following unconstrained formulation, {\it i.e.,}
               \bea
      F(\qx) &= &f(\qx) + \mu h^2(\qx) + \mu (\max(0,g(\qx))^2.
        \eea
        The main drawback about the penalty method is how to choose the penalty factors properly. If it is too large, it will cause suboptimality and numerical issues to the QUBO solvers while if it is too small, it will cause feasibility issues.

    \item  Augmented Lagrangian method is an alternative classical approach to the penalty method  \cite{nonlinear-book}.  Compared to the penalty method, it adds an additional Lagrange multiplier term to the objective function. For an equality-constrained problem, the augmented Lagrangian is expressed as
        \bea\label{eq:A0}
         \tilde F(\qx) & = & f(\qx) +  \lambda_h h(\qx) +  \frac{\mu}{2} (h(\qx) )^2 \notag,
         \eea
          and for a problem with both equality and inequality constraints, the augmented Lagrangian is expressed as
         \bea\label{eq:AL1}
         \tilde F(\qx) & = & f(\qx) +  \lambda_h h(\qx) +  \frac{\mu}{2} (h(\qx) )^2 \notag \\
          &+& \frac{1}{2\mu}\left(\max(0,\lambda_g + \mu g(\qx)))^2-\lambda_g^2\right),
         \eea
         where $\lambda_h$ and $\lambda_g$ are the Lagrange multipliers.
        In this way, instead of setting the  Lagrange multipliers and the penalty factors manually, the augmented Lagrangian method employs an iterative  solution  to update the original variables, the  Lagrange multipliers and the penalty factors. The iterative algorithm is briefly summarized in Algorithm 1 \cite{nonlinear-book}.

    \begin{algorithm}
\caption{To solve the constrained problem \eqref{prob0}}\label{alg:cap}
\begin{algorithmic}[1]
\Require \mbox{Initial~} $\lambda_h, \lambda_g,\mu$ \mbox{~and increasing factor~} $\rho$.
\Ensure   $\qx$,  \mbox{~and updated~} $\lambda_h, \lambda_g,\mu$ 
\Repeat
\State $\qx \leftarrow \arg\min_{\qx}  \tilde F(x)$,  where $\tilde F(x)$ is defined in \eqref{eq:AL1}.
\State $\lambda_h \gets \lambda_h + \mu h(\qx)$.
\State $\lambda_g \gets \max(0,\lambda_g + \mu g(\qx))$.
\State $\mu \gets \rho\mu$.
\Until{stopping criteria are satisfied.}
\end{algorithmic}
\end{algorithm}

Despite the theoretical interest of the above approaches, both techniques are not scalable and have limited performance, mainly due to their inefficiency to handle inequality constraints. Specifically, they are characterized by the following limitations:
\begin{enumerate}
\item The $\max(\cdot)$ operation required when dealing with the inequality constraint is not supported by standard QUBO solvers; existing commercial QUBO solvers can only handle linear and quadratic terms.
\item One way to overcome the above limitation is to introduce a slack variable  to convert the inequality constraint to an equality one, i.e., $g(\qx) + z^2 =0$. However, the slack variable $z$ will significantly increase the number of binary variables, because it is   a continuous variable and needs to be quantized using binary variables to be optimized. This will in turn require a large number of bits.
\item The augmented Lagrangian function in Line 2 of Algorithm 1 may result in higher-order terms than quadratic ones, even the original constraint function $g(x)$ is quadratic. This is because of the square function applied to the $\max(\cdot)$ operation.
\end{enumerate}

\end{itemize}

\section{The proposed HOBO Algorithm}
 In this section, we propose a modified augmented Lagrangian algorithm to solve Problem \eqref{prob0}, addressing the limitations of existing methods discussed earlier. Specifically, the proposed algorithm can handle both equality and inequality constraints without introducing slack variables. As previously noted, the augmented Lagrangian approach may lead to higher-order terms. Therefore, we begin by presenting an iterative algorithm for solving the unconstrained HOBO problem, followed by the introduction of the modified augmented Lagrangian algorithm.

\subsection{Unconstrained HOBO problem}
   QUBO solvers cannot tackle the HOBO problems directly, and most works in the literature resort to quadratization, which requires significantly more auxiliary variables and thus introduces higher complexity \cite{quadratization}. The details of the quadratization method is provided in the Appendix.

   In this section, we begin with designing an iterative algorithm to solve unconstrained HOBO problems which does not need any auxiliary variables. Suppose the objective function is $f(\qx)$ which is a higher-order polynomial. The main idea is to approximate $f(\qx)$  by its second-order Taylor expansion at a point $\qx_0$. The second-order approximation can be solved by QUBO solvers which leads to a solution $\tilde\qx$ and $\tilde \qx$ is in general a better solution than $\qx_0$. We continue to find the next Taylor approximation at the point $\tilde\qx$. In this way, we can minimize the approximate quadratic functions at each iteration until convergence. Below we introduce the details of the iterative algorithm design.

    Suppose the gradient vector and Hessian matrix of the original function is defined as
 \be
    \qg = \frac{\partial f(\qx)}{\partial \qx}, \qH_{i,j} = \frac{\partial^2 f(\qx)}{\partial x_i \partial x_j}.
 \ee
 Then the second-order Taylor approximation of $f(\qx)$ expanded at the point $\qx_0$ is expressed as
 \be\label{eqn:P:appro}
    \tilde f_{\qx_0}(\qx) = f(\qx_0)   +  \qg_{\qx_0}^T  (\qx-\qx_0)+ (\qx-\qx_0)^T \qH_{\qx_0} (\qx-\qx_0).
 \ee

   We can solve the approximate problem \eqref{eqn:P:appro} using a QUBO solver to obtain the solution $\tilde \qx$. To improve the solution $\tilde \qx$, we will evaluate whether it is a better solution than $\qx_0$, \textit{i.e.}, whether $f(\tilde \qx)< f(\qx_0)$ holds. If it holds, we perform the update $\qx_0=\tilde \qx$. If it does not hold, we will employ a greedy bit flipping method to modify and improve $\tilde \qx$. To be specific, we will change one bit of $\tilde \qx$ each time, \textit{i.e.}, $0\rightarrow 1$ or $1 \rightarrow 0$. If such a change results in a better solution, we keep the change (flip the original bit) and continue to evaluate the next bit flip; otherwise the original bit remains unchanged. This will increase the chance that from one iteration to the other, we can find a better solution, and therefore the resulting objective function is monotonically decreasing as iteration goes and the algorithm converges to a local or global optimum point.

\begin{figure}
	\centering
	\includegraphics[width=0.8\linewidth]{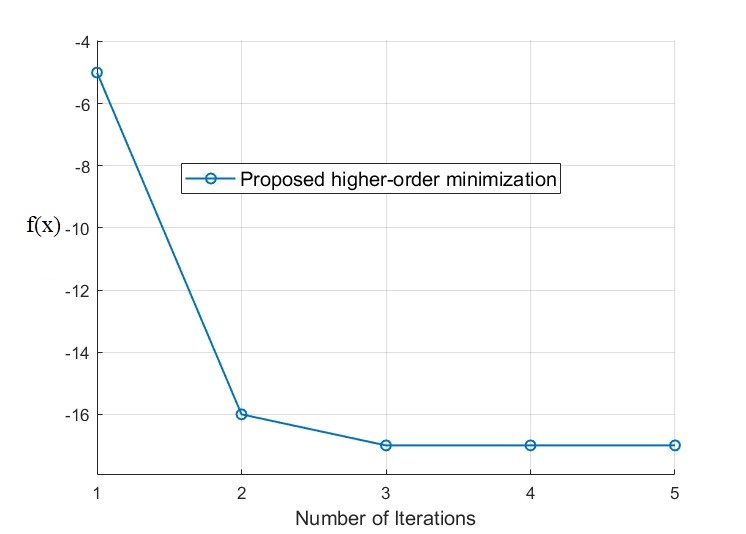}
	\vspace{-0.3cm}
	\caption{Convergence for solving a higher-order binary optimization problem using Algorithm 2.}\label{converg_approx}
\end{figure}

 The proposed algorithm to solve the unconstrained high-order problem $\min_{\qx} f(\qx)$ is summarized in Algorithm 2.
 It usually converges very fast after a few iterations. The convergence curve for a specific higher-order polynomial function $f(\qx)\triangleq  F(\qx, \lambda,\mu,v)$ with parameters $N=14,\lambda = 3, \mu = 5.5, v=1$, is given in Fig. \ref{converg_approx}, where $F(\qx, \lambda,\mu,v)$ is  defined later in \eqref{AL2}.

    \begin{algorithm}
\caption{To solve an unconstrained high-order problem with objective function $f(\qx)$ using a QUBO solver}\label{alg:cap}
\begin{algorithmic}[1]
\Require ~~An initial solution $\qx_0$.
\Ensure   An approximate solution $\qx^*$ that minimizes $f(\qx)$.
\Repeat
\State Construct the second-order Taylor approximation of $f(\qx)$ at $\qx_0$, which is  $\tilde f_{\qx_0}(\qx)$ defined in \eqref{eqn:P:appro}.
\State Solve $\tilde \qx = \arg\min_{\qx}    \tilde f_{\qx_0}(\qx)$ using a QUBO solver.
\State If $f(\tilde \qx)> f(\qx_0)$
\State ~~~~Perform bit flipping on the obtained solution $\tilde \qx$.
\State Update $\qx_0=\tilde \qx$.
\Until{Convergence.}
\State The approximate solution is given by $\qx^*=\tilde \qx$.
\end{algorithmic}
\end{algorithm}

\subsection{The modified augmented Lagrangian algorithm}
Having the Algorithm 2 to solve an unconstrained HOBO problem, we are ready to design a modified augmented Lagrangian algorithm to solve constrained binary optimization problems. For simplicity, we only consider the binary optimization problem with an inequality constraint, written as
 \bea\label{prob1}
    \min_{\qx\in\{0, 1\} \mbox{~or~} \{-1,1\}} && f(\qx) \notag \\
    \mbox{s.t.} && g(\qx)\le 0.
 \eea
 By introducing a slack variable $z$ to convert the inequality constraint to an equality one $g(\qx)+z^2=0$, we can formulate the following augmented Lagrangian, {\it i.e.,}
            \be\label{eqn:ag}
         F(\qx,z)  =  f(\qx) +\lambda_g (g(\qx)+ z^2) + \frac{\mu}{2} (g(\qx) + z^2)^2,
         \ee
where $z^2$ is chosen to minimize $F(\qx,z)$. By defining $v=z^2$, we have $F(\qx,v) = \frac{\mu}{2} v^2 + (\mu g(\qx) + \lambda_g) v +  \lambda_g  g(\qx) + f(\qx)  + \frac{\mu g^2(\qx)}{2}$ which is a quadratic function about $v$.

 The optimal $v$ is expressed as
 \be
  v=
\begin{cases}
    -\left[g(\qx) + \frac{\lambda_g}{\mu}\right],\,& \text{if } -\left[g(\qx) + \frac{\lambda_g}{\mu}\right]\ge 0;\\
    0, & \text{otherwise,}
\end{cases}
\ee
or more compactly
\be
v = \max\left(0,  -\left[g(\qx) + \frac{\lambda_g}{\mu}\right]\right).
\ee

Then the multiplier can be updated as
\be
\lambda_g = \lambda_g + \mu (g(\qx) + v) = \max(0,\lambda_g + \mu g(\qx)).
\ee

With the slack variable $v$, we can avoid the $\max(\cdot)$ operation in Algorithm 1, and therefore propose  the following modified augmented Lagrangian method,   which is summarized in Algorithm 3.
    \begin{algorithm}
\caption{To solve the constrained problem \eqref{prob1} using a QUBO solver}\label{alg:cap}
\begin{algorithmic}[1]
\Require \mbox{Initial~} $\lambda_g,\mu$ \mbox{~and the increasing factor~} $\rho$.
\Ensure   $\qx$   
\Repeat
\State Construct  the  augmented Lagrangian function $F(\qx,v)  =     f(\qx) +\lambda_g (g(\qx)+ v) + \frac{\mu}{2} (g(\qx) + v)^2 $.
\State Apply Algorithm 2 to solve \\  ~~~~~~ $\qx = \arg\min_{\qx}       F(\qx,v)$ using a QUBO solver.
\State Update $v = \max\left(0,  -\left[g(\qx) + \frac{\lambda_g}{\mu}\right]\right)$.
\State Update $\lambda_g = \max(0,\lambda_g + \mu g(\qx))$.
\State Update $\mu = \rho\mu$.
\Until{stopping criteria are satisfied.}
\end{algorithmic}
\end{algorithm}
The stopping criteria could include i) the maximum number of iterations is reached; ii) the original objective function does not change for a given number of iterations. Algorithm 3 can be extended to deal with multiple inequality and equality constraints.
The convergence of Algorithm 3 to a local optimum follows directly from Proposition 4.2.1 and Proposition 4.2.2 in \cite{nonlinear-book}.

 We also  emphasize that Algorithm 3 does not guarantee the global optimality. However, it offers an efficient approach for solving large-scale constrained HOBO problems by leveraging QUBO solvers.

\section{A case study: RIS optimization for simultaneous information and power transfer}\label{example}

To demonstrate the efficiency of the proposed method and its application interest in wireless communications systems, without loss of generality, a RIS-aided SWIPT system is considered as a case study. The emerging RIS technology enables on-demand reconfiguration of the wireless channel by leveraging passive reflecting elements that induce tunable phase shifts to the incident radio signal \cite{NASI}.  SWIPT refers to the co-engineering/co-design of radio-frequency signals to convey simultaneously information to conventional receivers as well as energy to low-power devices \cite{PSO}. In our example, a RIS with binary phase shift resolution is used to satisfy SWIPT objectives; the RIS phase shift vector is optimized to maximize information transfer, while satisfying a minimum energy harvesting (EH) requirement.

\subsection{System Model}

We consider a basic SWIPT system where a single-antenna transmitter serves both an information receiver and an EH receiver, both with a single-antenna. Direct links between the transmitter and the receivers do not exist (due to obstacles and/or strong shadowing/attenuation) and connectivity is facilitated by a passive RIS with $N$ reflecting elements. Due to hardware and power constraints, the phase shifts of the RIS have binary resolution and take values in the set $\{0,\pi\}$.

\begin{figure}
	\centering
	\includegraphics[width=\linewidth]{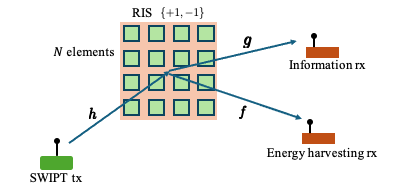}
	\vspace{-0.3cm}
	\caption{The system model; a single SWIPT transmitter conveys information to an information receiver and power to an energy harvesting receiver through an IRS with $1$-bit phase resolution. }\label{model}
\end{figure}

Suppose the wireless channel between the transmitter and RIS is denoted by $\qh\in \mathcal{C}^{N\times 1}$, while the channel from the RIS to the information and energy receivers are $\qg\in \mathcal{C}^{N\times 1}$ and $\qf\in \mathcal{C}^{N\times 1}$, respectively. The RIS phase shift vector is denoted by $\qx\in \mathcal{R}^{N\times 1}$ and each of its elements takes binary values $\{-1,1\}$ ({\it i.e.,} $\exp(j\pi)=-1$ corresponds to a phase shift $\pi$ and $\exp(j0)=+1$ corresponds to a phase shift $0$). Fig. \ref{model} schematically presents the system model.

Assume the transmit signal is $s$, then the received signals at the information and energy harvesting receivers, respectively, can be written as
\bea
    y_I &=& \qg^T \mbox{Diag}(\qx) \qh s + n_I, \\
    y_E& =& \qf^T \mbox{Diag}(\qx) \qh s + n_E,
\eea
with received power
 \bea
   P_I & = & \qx^T \qR \qx, \\
   P_E & =& \qx^T \qJ \qx,
 \eea
where $n_I$, $n_E\sim \mathcal{CN}(0,N_0)$ denote the additive white Gaussian noise at the information and energy receivers respectively, with variance $N_0$, and
\bea
    \qR &=& P\mbox{Re}\left(\mbox{Diag}(\qh^\dag)\qg\qg^\dag\mbox{Diag}(\qh)\right)\\
    \qJ &=& P\mbox{Re}\left(\mbox{Diag}(\qh^\dag)\qf\qf^\dag\mbox{Diag}(\qh)\right),
\eea
where $P$ is the transmit power.

The received signal to noise ratio (SNR) for the information receiver is expressed as
\be\label{obj:SNR}
    \Gamma = \frac{P_I}{N_0} = \frac{ \qx^T \qR \qx}{N_0}.
\ee

For the energy harvesting process, we adopt a nonlinear (quadratic polynomial) energy harvesting model \cite{BLE}. The harvested energy at the energy receiver (by ignoring the noise power) can be written as
\be
    EH = a_1 (\qx^T \qJ \qx) ^2 + a_2(\qx^T \qJ \qx) + a_3,
\ee
where $a_1,a_2,a_3$ are model parameters which can be obtained directly from harvesters’ data using standard convex optimization fitting methods.

\subsection{Problem Formulation}
 The problem of interest is to maximize the received SNR of the information receiver while satisfying the minimum energy $\delta$ that is harvested by the EH receiver. Mathematically it is formulated as
\bea\label{prob1}
    \min_{\qx\in \{-1,1\}^N} && -\qx^T \qR\qx \notag\\
    \mbox{s.t.} &&   a_1 (\qx^T \qJ \qx) ^2 + a_2(\qx^T \qJ \qx) + a_3 \ge \delta.
\eea

Note that the constraint is in a quadratic form of the input power $\qx^T \qJ \qx$, so it can be simplified and we can obtain a new equivalent problem formulation
\bea\label{prob2}
    \min_{\qx\in \{-1,1\}^N} && -\qx^T \qR\qx, \notag\\
    \mbox{s.t.} &&  \qx^T \qJ \qx  \ge c\triangleq \frac{-a_2 + \sqrt{4 a_1 (a_3 -\delta)}}{2a_1},
\eea
where \eqref{prob2} is an inequality-constrained  quadratic optimization problem and is difficult to solve in general. Next we will show how it can be solved by using our proposed Algorithm 3.

\subsection{Algorithmic  Solution}
To apply Algorithm 3 to solve \eqref{prob1}, we first form the augmented Lagrangian of \eqref{prob2} as
\bea\label{AL2}
    F(\qx, \lambda,\mu,v) &=& -\qx^T \qR\qx + \lambda (c- \qx^T \qJ\qx+v) \notag\\
    &+& \frac{\mu}{2}(c- \qx^T \qJ\qx+ v)^2\notag\\
    &= & -\qx^T ( (\lambda+ \mu(c+v)) \qJ +\qR ) \qx  + \frac{\mu}{2}(\qx^T \qJ\qx)^2 \notag\\
     &+&   \frac{\mu}{2}(c+v)^2 + \lambda  c.
\eea
Clearly $F(\qx, \lambda,\mu,v)$ is a higher order polynomial which consists of quartic terms in addition to quadratic terms. The objective function $F(\qx, \lambda,\mu,v)$ can be minimized with regard to $\qx$  by using Algorithm 2, which can be solved by a QUBO solver subsequently. With the above formulation, we can apply Algorithm 3 to solve the RIS phase shift optimization problem in \eqref{prob1}.

\section{Numerical results}

In this section, numerical simulations are carried-out to evaluate the performance  of the proposed algorithm by using various  QUBO solvers;  the problem of RIS phase shift optimization is used as an illustrative example.

The simulation setup follows the description of Section \ref{example}. Specifically, the distance between the transmitter and the RIS is $3$ m with a direction angle of $\frac{\pi}{4}$.  The information and energy receivers are uniformly  located within a distance ranging in $(5, 30)$ m and $(1, 2.5)$ m from the RIS, respectively.  The operating frequency is set as $f=915$ MHz and it is assumed that the antenna gains  at the transmitter and receivers are $8$ dBi and $0$ dBi, respectively. The path-loss attenuation for all channels is obtained by using the Friis equation with a reference distance $1$ m and the path-loss propagation exponent $2$.

Because of the relatively short distance between the RIS and the transmitter/receivers, the line-of-sight (LOS) signal dominates the transmission and therefore the Rician fading is used to model the small-scale fading for all channels. Hence, $\mathbf{h}$ is composed of the  LOS  part $\mathbf{h}^{LOS}$ and the non-LOS part  $\mathbf{h}^{NLOS}$ as follows
\begin{align}
	\mathbf{h} &= \sqrt{\frac{K}{1+K}} \mathbf{h}^{LOS}_k + \sqrt{\frac{1}{1+K}} \mathbf{h}^{NLOS}_k,
\end{align}
where $K=5$ dB is the Rician factor.  Rayleigh fading is adopted for the NLOS signal, $\mathbf{h}^{NLOS}_k\in\mathbb{C}^{N\times 1}$  and each of its elements is a   complex Gaussian random variable with zero mean and unit variance. In a similar way, we define the wireless channels $\mathbf{g}$ and $\mathbf{f}$.

Unless otherwise specified, it is further assumed that the transmit power is $40$ dBm and the noise power is assumed to be $-60$ dBm.
The fitting parameters of the   energy harvesting model  are $a_1=-1.2006 \times 10^{-4}, a_2 =  0.6734, a_3 = -3.5988$ by using the data in \cite{SONG} based on real-world measurements using curve fitting tools; the input and output power is in microwatt.

In the proposed modified augmented Lagrangian  Algorithm 3, we employ the    dSB  introduced in Section III and implemented by the open-source Python package \cite{SB_implementation} together with the proposed iterative Algorithm 2 to solve the higher-order optimization. This method is labelled as `SB'.

For comparison, we use the exhaustive search method (when possible) as the optimal solution and performance upper bound by searching all $2^N$ solutions for $N\le 24$. The random phase shift is used as a performance lower bound by generating $5,000$ random phase shift solutions and selecting the best one.

In addition, we also consider the quadratization method \cite{quadratization} instead of Algorithm 2 to solve the higher-order problem in Algorithm 3 using the following solvers, and they serve as performance benchmarks for comparison:
\begin{enumerate}
  \item Simulated annealing (SA). It is a widely used classical algorithm for solving unconstrained optimization problems including QUBO; we use the implementation from D-WAVE \cite{WAVE}.
  \item Amplify.  It uses Amplify's Annealing Engine (with free license) introduced in Section III  which can obtain high-quality solutions for large-scale QUBO problems. Note that for this method, we need to access Amplify's cloud server and this incurs a significant  amount of computation latency.
  \item Gurobi. It is a commercial numerical solver for solving a wide range of optimization problems \cite{Gurobi}; we use the academic version of it to solve the QUBO problem in the quadratization.
   \item Penalty method. We will also include the comparison with a conventional penalty method which solves the following problem for a large penalty factor $\lambda$, {\it i.e.,}
\bea\label{prob3}
    \min_{\qx\in \{-1,1\}^N} && \!\!\!\!\!\!-\qx^T \qR\qx  + \lambda \max(c-\qx^T \qJ \qx,0  )^2.
\eea
It is a higher-order binary optimization problem but it can be solved by the SA algorithm. If solving the above problem does not return a feasible solution, then we increase the penalty factor by $10$ times and   repeat solving the problem \eqref{prob3}. It is well known that the penalty method  has the inherent drawback that  as the penalty coefficient grows and the quadratic  coefficients are large, the associated  QUBO problem becomes ill-conditioned, and this may cause numeric errors and slow convergence.
\end{enumerate}

For Algorithm 3, we set the initial values $\lambda =3, \mu = 5.5, \rho=1.1$. The minimum and maximum numbers of iterations are $10$ and $50$, respectively. We choose the best solution after the algorithm is terminated.

\subsection{The numbers of variables and polynomial terms}
We first show the number of variables as well as the number of the linear and quadratic terms after quadratization as the number of RIS elements $N$ varies. We can see that as $N$ increases, the number of variables and the number of quadratic terms quickly increase.  This increase is usually problem dependent. For the RIS optimization problem, from \eqref{AL2}, we can see that  the number of quartic terms has the order of $\mathcal{O}(N^2)$, and from the Appendix we know that  the total number of variables after quadratization also has the order of $\mathcal{O}(N^2)$. The number of original quadratic terms in \eqref{AL2} has the order of $\mathcal{O}(N^3)$, and additional quadratic terms due to quadratization   has the order of $\mathcal{O}(N^2)$. Therefore, the number of overall quadratic terms still has the order of $\mathcal{O}(N^3)$.  This can be verified by the results in Table \ref{tab:var}.  Compared to the proposed Algorithm 2 that does not need extra variables to solve the higher-order problem,  quadratization needs to solve significantly more binary variables. Therefore, it increases the complexity and is not scalable for large $N$, as evidenced later.

 \begin{table}[h]
 \scriptsize
\centering
\caption{The numbers of Ising variables, linear and quadratic terms vs  the number of RIS elements, $N$.}
\label{tab:var}
\begin{tabular}{ |l|l|l|l|l|l|l|l| }
\hline
      $N$                    & 10    & 12      & 16      & 20 & 24 & 32& 36 \\ \hline
  $\#$ of variables         &  92   & 144     &  256    & 400   & 576 &1024 & 1296 \\ \hline
  $\#$ of linear terms        &  92   & 144     &  256    &  400  & 576  &1024 & 1296  \\ \hline
$\#$ of quadratic terms      &    472   &  891  &   2540  &  5985  &  12282  & 38936 &62685  \\ \hline
\end{tabular}
\end{table}

\subsection{The effect of the number of RIS elements $N$}
We then evaluate the feasibility performance  of the exhaustive search and random search in Table \ref{tab:fea:N}  assuming the EH threshold ($\delta$) is $500$ microwatt. The results are obtained by randomly generating $10,000$ channel realizations.  We can see that as the number of RIS elements $N$ increases, the feasibility improves for the exhaustive search method. For the random phase shift algorithm, the feasibility remains low which means it can hardly satisfy the EH requirements without optimization.

\begin{table}[h]
\centering
\caption{The probability of feasibility vs  the number of RIS elements, $N$, with  $\delta=500$.}
\label{tab:fea:N}
\begin{tabular}{|l|l|l|l|l|l|  }
\hline
      $N$         & 10      & 14         & 16 & 20   & 24 \\ \hline
  Exhaustive      &   0.0696   & 0.3174         &   0.4532  & 0.7073& 0.9218   \\ \hline
 Random &      0.0118   &  0.0320  &  0.0420    &  0.0814  & 0.1361 \\ \hline
\end{tabular}
\end{table}

Next we proceed with our evaluation by choosing $100$ channels that are feasible when using the exhaustive search method. Specifically, Table \ref{tab:fea2:N} shows the feasibility results  for these $100$ channels, by adopting the proposed Algorithm 3 and quadratization with various QUBO solvers. It can be seen that all solvers achieve similar and high  feasibility probabilities. This observation has been expected since as the number of RIS elements increases, it becomes easier for the system to satisfy the EH requirement.
\begin{table}[h]
\centering
\caption{The probability of feasibility vs  the number of RIS elements, $N$ with  $\delta=500$.}
\label{tab:fea2:N}
\begin{tabular}{ |l|l|l|l|l|l|  }
\hline
      $N$           & 10      & 14       & 16 & 20   & 24 \\ \hline
 SA    &    0.99  & 1          &  1  &1& 1    \\ \hline
  Penalty &   1  &   1   & 1          &  1  &1    \\ \hline
 \textbf{SB}     &   0.97  & 1       &  1  &1& 1   \\ \hline
 Amplify     &    0.97  & 0.99          &  1  &1& 1   \\ \hline
 Gurobi    &    0.98   &  0.97 &  1 &  1    &  1   \\ \hline
\end{tabular}
\end{table}

\begin{figure}[]
	\centering
	\includegraphics[width=\linewidth]{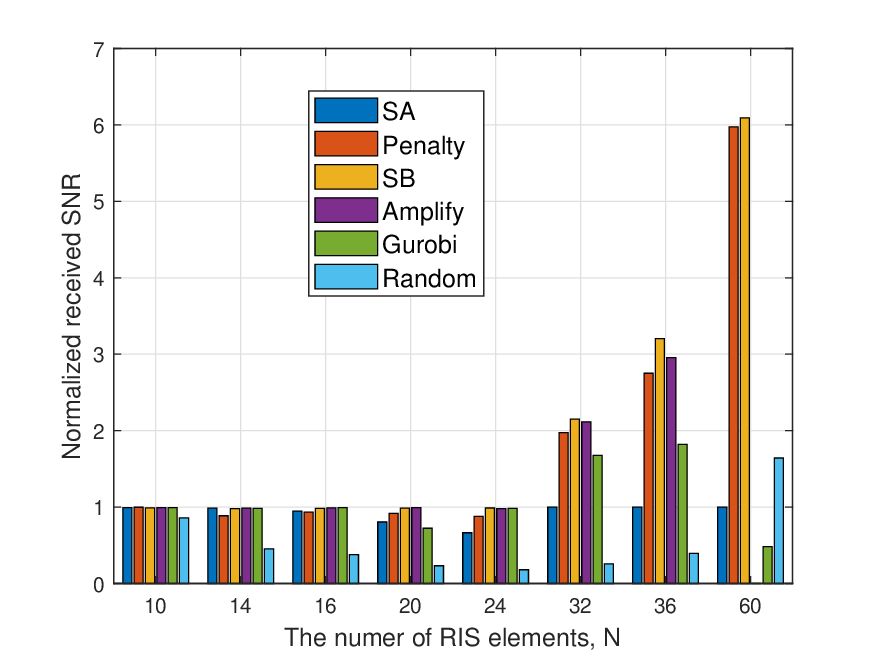}
	\vspace{-0.3cm}
	\caption{Visualization of results in Table \ref{tab:SINR:N}.}\label{fig2}
\end{figure}

In Table \ref{tab:SINR:N}, we evaluate the achievable relative SNR performance when using different QUBO solvers and random search. The results are also visualized in Fig. \ref{fig2}.  The SNR results are normalized with respect to the exhaustive search method when $N\le 24$, and by using feasible channel realizations for each solver.
For scenarios with $N>24$, the results are normalized with respect to that of the SA method, since exhaustive search solution cannot be computed in this case. Because not all channels lead to
feasible solutions as shown in Table \ref{tab:fea2:N}, for each scheme, we only choose those feasible channels to normalize its performance over
the exhaustive search or SA method.  It can be seen that the performance of the random search method becomes poorer as the number of RIS elements increases. This is because the search space becomes huge and $5,000$ random searches are not sufficient to get  satisfactory performance. The SA solver achieves similar performance as SB and Amplify heuristics for small $N$ but  when $N\ge 20$, its performance degrades significantly  because it cannot handle the large number of variables. A similar trend can be observed for the penalty method.
 The proposed SB method and Amplify both achieve similar and near optimal performance, but our proposed method requires to solve only $N$ variables at each iteration, as opposed to $N^2$ for Amplify and outperforms Amplify when $N\ge 32$. Gurobi achieves similar performance to  Amplify only when $N$ is small, which indicates it cannot obtain stable performance as the number of variables become large. This becomes more obvious for large RIS topologies with $N\ge 32$, where clear gaps between the SB  and Amplify/Gurobi solutions can be observed. When $N=60$, the performance of Gurobi is even worse than that of the random search. It takes excess time for Amplify and Gurobi to solve the problem using quadratization due to the large number of extra variables so their performance is not shown for $N=60$.

\begin{table}[h]
\tiny
\centering
\caption{The effect of the number of RIS elements, $N$ on SNR. Performance normalized with respect to the exhaustive search method. For $N\ge32$, it is normalized with respect to the SA performance; $\delta=500$.}
\label{tab:SINR:N}
\begin{tabular}{|l|l|l|l|l|l|l|l|l|}
\hline
      $N$          & 10      & 14         & 16 & 20& 24& 32& 36 &60 \\ \hline
 SA    &  0.9941  & 0.9855  &  0.9461  &  0.8048  &  0.6642 &1 &1 & 1\\ \hline
 Penalty    & 0.9976  &  0.8866  &  0.9349  &  0.9185 &   0.8777 &1.9739  &  2.7510 & 5.9743\\ \hline
  \textbf{SB}   & 0.9902   & 0.9806  & 0.9820  & 0.9849   & 0.9879 & 2.1518  &  3.2034&  6.0916 \\ \hline
 Amplify    &   0.9929 &    0.9855 &    0.9904 &    0.9932&  0.9788&  2.1137& 2.9544 &  -  \\ \hline
 Gurobi  &    0.9929 &    0.9832 &    0.9915 &    0.7231 &  0.9835 & 1.6759&1.8210& 0.4823  \\ \hline
 Random   &    0.8598  &   0.4542   &    0.3769 &     0.2323 &    0.1797 & 0.2561&0.3931& 1.6418 \\ \hline
\end{tabular}
\end{table}

\begin{figure}[]
	\centering
	\includegraphics[width=\linewidth]{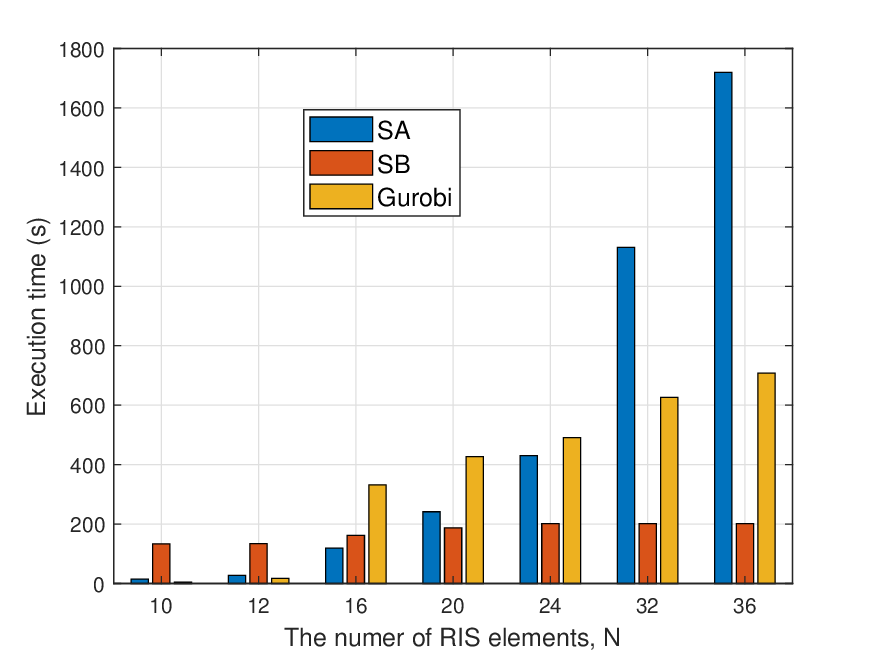}
	\vspace{-0.3cm}
	\caption{Execution time versus the number of RIS elements,  $\delta=500$.}\label{fig_time}
\end{figure}

\begin{figure}[]
	\centering
	\includegraphics[width=\linewidth]{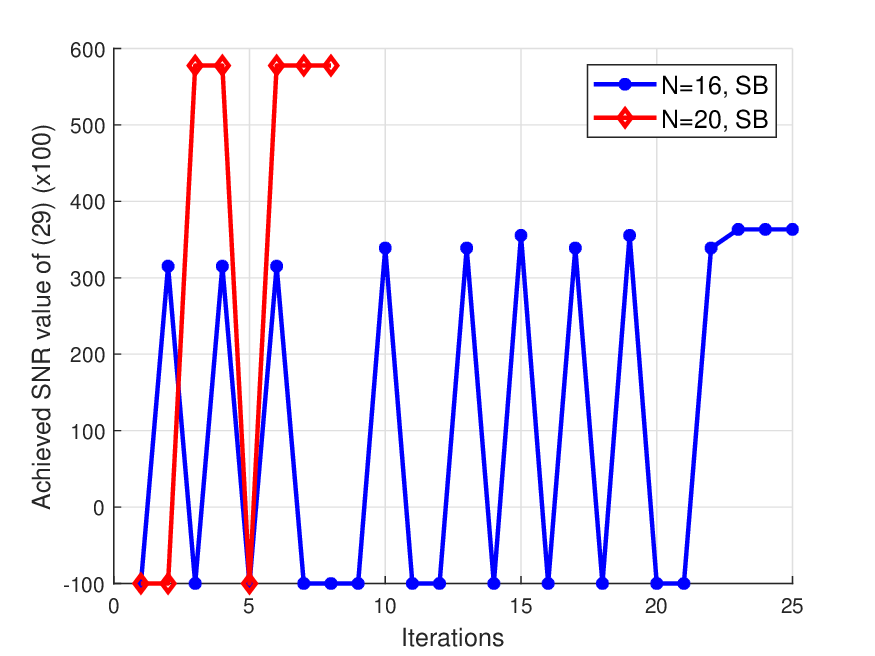}
	\vspace{-0.3cm}
	\caption{Convergence of the proposed  method (SB),  $\delta=500$.}\label{fig_conv}
\end{figure}

 Next we evaluate the complexity of various QUBO solvers using the proposed algorithm. Because different solvers use different computing resources and theoretical characterization of their complexities is not available, it is not straightforward to make a fair comparison. Instead,  we depict the execution time versus the number of RIS elements $N$ in Fig. \ref{fig_time}.  It is measured as the time difference before and after calling the corresponding QUBO solvers.  Because Amplify requires access to the cloud server, it is not included in the comparison. As can be seen, as $N$ increases,  all algorithms require more time to complete the computation. The required time by the SA algorithm  increases the fastest and this is because the complexity of SA is $\mathcal{O}(e^N)$ \cite{complexity}. As $N$ increases, SB requires the least time and remains stable.   The convergence behaviour of the proposed SB method is illustrated in Fig. \ref{fig_conv} for N=16 and N=20. The figure plots the objective function defined in equation \eqref{obj:SNR}, which corresponds to the achieved SNR value, versus the number of iterations. If the solution in a given iteration is infeasible, the objective value is set to –100 to indicate infeasibility.

  Note that time-to-solution (TTS) is a useful metric that measures the time required to obtain the optimal solution with a certain probability, typically specified as 0.99 \cite{TTS}. To calculate the TTS, a solver needs to obtain at least a few samples of the optimal solution to calculate its probability of occurrence.   However, for the Amplify and SB solvers, we cannot obtain multiple solutions and it is not practical to get the optimal solution through exhaustive search for large $N$, and therefore we do not adopt it as a performance measure. This is a limitation of the QUBO solvers used in this paper, and TTS can still be adopted  for solvers that can produce multiple solutions such as D-Wave's quantum annealer.

\subsection{The effect of  the EH requirement $\delta$}
In this subsection, we investigate the effect of the   EH requirement $\delta$ when $N=20$  as $\delta$ varies from $200$ to $800$ microwatt. First, we evaluate the feasibility performance  in Table \ref{tab:fea2:EH} for different algorithms.
It is observed that when the EH requirement is high, it is more difficult to satisfy even for the exhaustive search method; for the random search method, the feasibility of satisfying the EH constraint remains low as $\delta$ increases. We can see that proposed SB algorithm and other QUBO solvers achieve the same feasibility probabilities as the exhaustive search method which are all reduced as the EH requirement becomes more demanding.
\begin{table}[h]
\centering
\caption{The probability of feasibility vs  the EH constraint, $\delta$ in microwatt, $N=20$.}
\label{tab:fea2:EH}
\begin{tabular}{|l|l|l|l|l|l|  }
\hline
      $N$ & 200        & 300           & 500  & 700   & 800  \\ \hline
 SA &   1  &   1   & 1    &  0.90 &0.76 \\ \hline
 Penalty    &   1  &   1   & 1    & 0.90 &0.76   \\ \hline
  \textbf{SB} &  1  &   1   & 1    & 0.90 &0.76\\ \hline
 Amplify &  1  &   1   & 1    &  0.90 &0.76  \\ \hline
 Gurobi &  1  &   1   & 1    &  0.90 &0.76  \\ \hline
 Random & 1  &   1&   0.87&   0.30&  0.17  \\ \hline
\end{tabular}
\end{table}

\begin{figure}[]
	\centering
	\includegraphics[width=\linewidth]{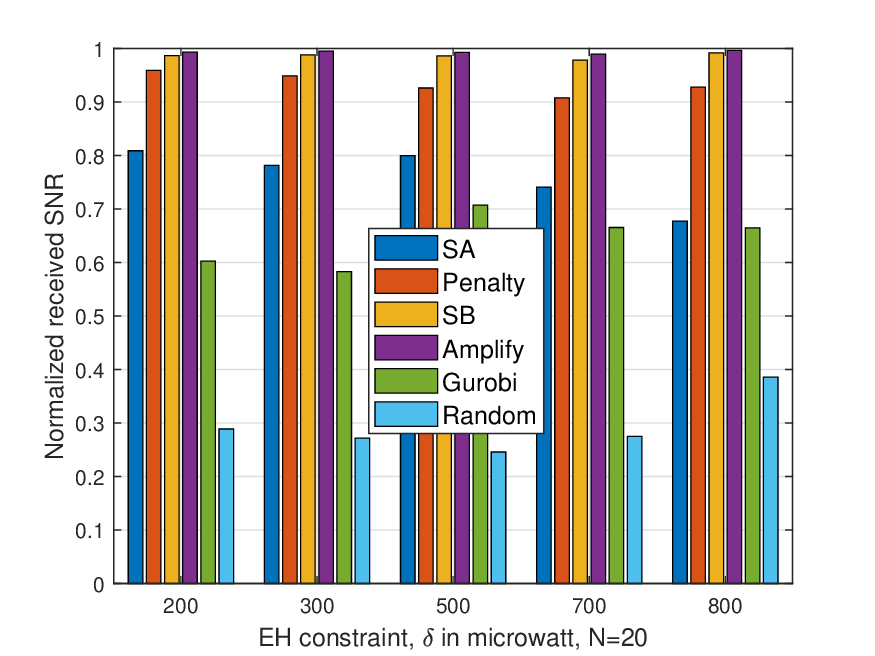}
	\vspace{-0.3cm}
	\caption{Visualization of results in Table \ref{tab:EH:N20}.}\label{fig4}
\end{figure}

In Table \ref{tab:EH:N20}, we show the  achievable relative SNR  of different schemes with varying EH requirements for the feasible channels (visualized in Fig. \ref{fig4}). The results have been normalized with respect to the exhaustive search method. We observe that the both the proposed SB method and Amplify   can achieve consistent and  near optimal performance and outperform the penalty method, while the proposed method has the lower complexity advantage over Amplify.  Although both SA and Gurobi are commercial solvers, their performance degrades significantly when $N$ is large, as explained by the results in Table I. This is primarily due to their inability to efficiently handle the large number of variables introduced by the quadratization method.  Similar observations  can be made via results  shown in Table \ref{tab:prob95:EH} regarding the probability that various methods achieve at least 95\% performance of the exhaustive search method.
 \begin{table}[h]
\centering
\caption{The effect of the EH constraint on SNR, $\delta$ in microwatt. Performance normalized by that of the exhaustive search  method, $N=20$.}
\label{tab:EH:N20}
\begin{tabular}{|l|l|l|l|l| l|}
\hline
      $\delta$ & 200        & 300           & 500  & 700   & 800 \\ \hline
SA & 0.7915  &  0.8050 &   0.8040  &  0.7429  &  0.6204\\ \hline
 Penalty &  0.9591    &  0.9487  &  0.9262   &  0.9076   & 0.9278 \\ \hline
    \textbf{SB}  &   0.9868  &  0.9879 &   0.9861 &   0.9783&    0.9920 \\ \hline
   Amplify  &   0.9933  &  0.9951 &   0.9926 &   0.9894  &  0.9965  \\ \hline
    Gurobi  & 0.6026   & 0.5830   & 0.7073&    0.6654  &  0.6647 \\ \hline
 Random  &   0.2888   &   0.2717   &   0.2458 &     0.2751 &     0.3859  \\ \hline
\end{tabular}
\end{table}

\begin{table}[h]
\centering
\caption{The probability of achieving 95\% performance  vs  the EH constraint, $\delta$ in microwatt, N=20.   }
\label{tab:prob95:EH}
\begin{tabular}{|l|l|l|l|l|l| }
\hline
      $\delta$  & 200          & 300          & 500   & 700 & 800 \\ \hline
 SA & 0.13  &    0.22&     0.33&     0.37&     0.22  \\ \hline
 Penalty & 0.68 &    0.64 &     0.54 &    0.53&     0.44  \\ \hline
   \textbf{SB}  &   0.89 &     0.89&     0.84&    0.73&    0.65   \\ \hline
   Amplify  &    0.92&   0.91&   0.92&   0.79&    0.68\\ \hline
   Gurobi  &   0.02 &   0.07 &    0.21 &  0.27&    0.15  \\ \hline
  Random  &  0    &  0  &    0  &     0  &    0    \\ \hline
\end{tabular}
\end{table}

\section{Conclusions}
In this paper, we have presented an  algorithmic   solution  to solve the HOBO
problem in wireless communications systems with inequality constraints using Ising machines. The proposed algorithm is based on the augmented Lagrangian method and
incorporates appropriate  techniques and efficient Ising machines to solve higher-order problems without introducing auxiliary variables. By using the example of optimizing the binary phase shifts in a RIS-assisted SWIPT network, we have demonstrated the effectiveness of the proposed algorithm. We observed that in general most conventional heuristics including simulated annealing experience severe performance degradation as the number of variables is large. Results verified that the proposed algorithm can still achieve near-optimal performance for large-scale HOBO problems, and therefore it serves as a promising alternative to classical heuristic algorithms and purely quantum optimization in the noisy intermediate-scale quantum era.  With the advancement of both conventional and quantum solvers, the proposed algorithm has the potential to empower future quantum systems and classical Ising machines to tackle large-scale optimization problems in wireless communications that remain intractable for current heuristic approaches.

\section*{Appendix: Quadratization}
 In the appendix, we  describe a quadratization method that converts a higher-order polynomials to a quadratic polynomial  \cite{quadratization} such that it can be handled by QUBO solvers.

 We first present the following results for a cubic term with regard to Boolean and Ising variables.
 \begin{lemma}
 For Boolean variables $x_1, x_2,x_3$,  minimization of a cubic term  $x_1 x_2 x_3$ can be equivalently written as
 \be
    \min_{x_1, x_2, x_3,y\in\{0,1\}}   y x_3 + M (3y + x_1x_2 -2x_1y - 2x_2y),
 \ee
 where $M$ is a large positive constant and  $y$ is an auxiliary Boolean variable.
 \end{lemma}

 Lemma 1 was first proposed in \cite{Rosenberg} and is also widely known as Rosenberg’s polynomial. It is derived by replacing the condition $y=x_1 x_2$ with a new penalty term. It can reduce the order of a high-order polynomial by one by introducing one extra variable.

 On the contrary, Lemma 1 does not work directly for Ising variables. Instead, we need the following Lemma 2 to reduce the order of a cubic term for Ising variables \cite{quadratization}, {\it i.e.,}
 \begin{lemma}
  For Ising variables $x_1, x_2,x_3$,  minimization of a cubic term  $x_1 x_2 x_3$ can be equivalently written as
 \bea
    \min_{x_1, x_2, x_3,y, d\in\{-1,1\}}   y x_3 + M (4 + x_1 + x_2 - y - 2d+ x_1x_2 \notag \\- x_1y - x_2y - 2x_1d - 2x_2 d + 2yd),
 \eea
 where $M$ is a large positive constant and  $y$ and $d$ are   auxiliary Ising variables.
 \end{lemma}
 Lemma 2 is derived by replacing the condition  $y=x_1 x_2$ with a new penalty term for the Ising variables, and it needs two extra variables to reduce the order of a high-order polynomial by one.

Boolean and Ising QUBO formulations are equivalent by a simple linear transformation. However,  note that it is advantageous to have separate  results for reducing the order for Boolean and Ising polynomials, respectively, because a sparse polynomial in one space is not necessarily sparse in the other. For instance, a single Ising term $\prod_{i=1}^N x_i, x_i \in \{-1,1\}$ can be converted to  $\prod_{i=1}^N (2 b_i-1),  b_i \in \{0,1\}$ which consists of $2^N$ terms in a Boolean space. Therefore, to reduce the order of $\prod_{i=1}^N x_i$, it is preferred to do it in the Ising space by using Lemma 2 rather than first converting it to the Boolean space and then use Lemma 1.

 The results in Lemma 1 and Lemma 2 serve as the foundation to reduce the order of high-order polynomials; by introducing new variables and repeating Lemmas 1 and  2, we can eventually reduce high-order polynomials to quadratic ones. This process is called quadratization.  An efficient  quadratization algorithm that can potentially
give significant savings in the number of auxiliary variables has been proposed in \cite{quadratization} and is summarized in Algorithm 4.

 Note that the number of auxiliary variables required for quadratization largely depends on the interactions among variables and the number of higher-order terms. As shown in Table I, dense problems typically require a large number of auxiliary variables. However, for large but sparse problems, quadratization may still be preferred due to its computational efficiency.

 \begin{algorithm}[htbp]
\caption{{\scshape Quadratization of a higher order Polynomial \cite{quadratization}}}
\label{fig:hobotoqubo1}
\begin{algorithmic}[1]

\Require{A higher-order binary polynomial to be minimized over $(x_1, x_2, \dots, x_n)$ where $x_i  \in \{0,1\}$ or $\{-1,1\}$.}

\Ensure{A QUBO equivalent to the given higher-order binary minimization problem.}

\State Sort the indices of variables according to values and create a data structure of (key, value) pairs where the key is the set of all quadratic terms appearing in higher-order binary polynomial and the value is the set of all monomials of degree at least $3$.

\While{all the keys are deleted}

    \State Select the key with the largest number of values and replace it with an auxiliary variable.

    \State Update the data structure by adding keys and values for the new variable.

    \State Delete all degree-3 terms that involve  the key.

    \State Delete the key if all the values of the key has been deleted.

    \State Store the variable and the quadratic term it substitutes in a map.

\EndWhile

\State Invoke the quadratic polynomial corresponding for each map of auxiliary variable and the quadratic term using Lemma 1 or Lemma 2.

\State {\bf Return} The QUBO equivalent to the given  higher-order binary polynomial.
\end{algorithmic}
\end{algorithm}

\end{document}